\documentclass[12pt]{article}

\def\d{\partial}

\def\d{\partial}

\def\({\left(}
\def\){\right)}
\def\[{\left[}
\def\]{\right]}

\def\({\left(}
\def\){\right)}

\def\d{\partial}

\def\beq{\begin{equation}}
\def\eeq{\end{equation}}
\def\bea{\begin{eqnarray}}
\def\eea{\end{eqnarray}}
\def\bq{\begin{quote}}
\def\eq{\end{quote}}

\def\d{\partial}

\def\({\left(}
\def\){\right)}

\def\g5{\gamma_5}

\def\gappeq{\mathrel{\rlap {\raise.5ex\hbox{$>$}}
{\lower.5ex\hbox{$\sim$}}}}

\def\lappeq{\mathrel{\rlap{\raise.5ex\hbox{$<$}}
{\lower.5ex\hbox{$\sim$}}}}

\def\Toprel#1\over#2{\mathrel{\mathop{#2}\limits^{#1}}}

\begin{document}

\pagestyle{empty}
\begin{flushright}
{\ttfamily hep-th/0304059}\\
{CPHT-RR 012.0303}\\ 

\end{flushright}
\vspace*{5mm}
\begin{center}
{ \Huge \Huge Modified Star-Products  \\
 \vspace*{5mm}Beyond the Large-$B$ Limit} \\
\vspace*{19mm}
{\large Pascal Grange} \\
\vspace{0.5cm}

 {\it Centre de physique th{\'e}orique de l'{\'E}cole polytechnique,\\
 \vspace*{0.2cm}route de Saclay, F-91128 Palaiseau Cedex}\\
\vspace{0.5cm} 
{\tt{ pascal.grange@cpht.polytechnique.fr}}\\

\vspace*{3.5cm}
{\bf ABSTRACT} \\ \end{center}
\vspace*{4mm}
\noindent
 
Derivative corrections to the Wess--Zumino couplings of open-string
effective actions are computed at all orders in derivatives, taking the open-string metric into
account. This leads to a set of deformed star-products beyond the
Seiberg--Witten limit, and allows to
reinterpret the couplings in terms of a deformed integration
prescription along a Wilson line in the non-commutative
set-up. Moreover, the recursive definition of the star-products
induces deformations of $U(1)$ non-commutative {\mbox{Yang--Mills}} theory.\\

\noindent{PACS-codes}:  11.15, 11.25.\\
\noindent{Keywords}: D-branes, effective action, non-commutativity, open strings.\\

\vspace*{0.5cm}
\noindent

\begin{flushleft} 
April 2003
\end{flushleft}

\vfill\eject

\setcounter{page}{1}
\pagestyle{plain}

\section{Introduction}

Non-commutative field
theory on the world-volume of a D-brane has been developed in a
peculiar limit where a large constant background $B$-field is turned
on~\cite{CH,SW}. This is called the Seiberg--Witten limit and amounts
to $\alpha'\rightarrow 0$ together with a scaling of the metric,
$g_{ij}\sim\alpha'^2$, while open-string parameters are kept fixed. Duality properties have been studied, and an explicit mapping
between ordinary and non-commutative gauge fields exhibited~\cite{Cornalba,L,OO}. This
inspiring correspondence has been successfully extended to the
Ramond--Ramond couplings in the Seiberg--Witten limit, leading to an infinite set of
derivative corrections~\cite{CSterms,LM,DMS}, since the corrections
are suppressed by powers of $\alpha'$ in the non-commutative set-up. The
series of corrections are expressed in terms of modified star-products
named $\ast_p$ (for integer $p$), that arise naturally from an
integration prescription along an open Wilson line. This prescription
originates from the requirement of gauge invariance of observables in
non-commutative field theory~\cite{Rey1,Rey2,Gross,star-trekI,MW,DT,ReyHouches}. Computations at tree level in presence of a
single Euclidean {\mbox{D9-brane}} in commutative string theory
provided successful checks~\cite{M, mezigue}, confirming that
these corrections are leading in the Seiberg--Witten limit. This suggested that the
correspondence could be extended by string computations beyond this limit. Mukhi and
Suryanarayana~\cite{MSbeyond} derived the first correction in terms of the
open-string metric to the coupling of quadratic order in the field
strength, and generalized it to all orders (in the metric) by using a
disk amplitude computed by Liu and Michelson~\cite{star-trekIII}. This led to a
deformation of the $\ast_2$-product by a differential operator $t$
constructed out of the open-string metric $G$: 
$$C^{(6)}(-k)\wedge\int dx\, \langle F\wedge F\rangle_{\ast_2}e ^{ikx}\mapsto
C^{(6)}(-k)\wedge\int dx\, \langle F\wedge F\rangle_{\ast_2(t)}e ^{ikx},$$
\noindent{where} the $\ast_2(t)$ has the expected Seiberg--Witten
limit
$$t~:=\alpha'\d G\d',\,\,\,\,\,\,\,\,\,\,\,a~:=\frac{\d\theta\d'}{2\pi}, $$
$$\ast_2(t)=\frac{\Gamma(1+2t)}{\Gamma(1-a+t)\Gamma(1+a+t)},$$
$$\ast_2(0)=\frac{\sin{\pi a}}{\pi a}=\ast_2.$$
The deformed star-product $\ast_{2}(t)$ received an interpretation
in terms of a deformed {\mbox{smearing}} prescription along an open Wilson
line, that parallels the one that had given rise to the
{\mbox{$\ast_2$-product}}, since
$$ \int dx\,\langle F\wedge F\rangle_{\ast_2(t)} e ^{ikx}=\int dx\int_0^1d\tau\, F(x) \ast(t)\wedge F(x+\theta k\tau)\ast
e ^{ikx}, $$
$${\mathrm{ where }}\,\,\, \ast(t)=\ast\times\frac{\ast_{2}(t)}{\ast_2}=\ast\times\frac{\Gamma(1-a)\Gamma(1+a)\Gamma(1+2t)}{\Gamma(1-a+t)\Gamma(1+a+t)}.$$
{\noindent In} this letter we shall 
derive the contribution of the open-string metric to the amplitude
$$S_{CS}+\Delta S_{CS}=\langle C|\exp\(-\frac{i}{2\pi\alpha'}\int d\sigma d\theta D\phi^\mu A_\mu(\phi)\)|B\rangle_R,$$
\noindent{where $\phi^\mu=X^\mu+\theta\psi^\mu$} denotes a superfield,
and $D$ a derivative in superspace. This will enable us to derive the prescription of~\cite{MSbeyond}  and to extend it to larger
orders in the field strength.\\

The recursive definition of the modified star-products allows to
address the question of the correct definition of gauge tranformation
  laws beyond the large-$B$ limit. We shall
work with a single Euclidean D9-brane in the description where the open-string
metric $G$ is defined by
$$  \(\frac{1}{g+2\pi\alpha' B}\)^{ij}=\frac{\theta ^{ij}}{2\pi\alpha'}+G^{ij}.$$  
\noindent{We} shall first work out the deformation of the star-products in
presence of the symmetric part of the two-point functions of
 world-sheet scalars,
 and then interpret the results in terms of a deformed gauge-invariant
 smearing
 prescription in the non-commutative set-up.

\section{ Taking the open-string metric into account}

\subsection{Quadratic order in the field strength }
As noticed in previous investigations of Ramond--Ramond couplings for
small field strength, we are instructed to compute couplings whose
order in the field strength is half of the degree as a differential
form. Writing the result in terms of the differential form $F$, one
has to 
make the substitution
$$\frac{1}{2}\psi_0^\mu\psi_0^\nu F_{\mu\nu}\mapsto -i\alpha'F.$$ 
\noindent{The} only role played by the fermions in our
computation will therefore be to provide us with the suitable
number of zero modes, in order to build the grading of the coupling. As we are dealing with the Ramond--Ramond sector, we may forget the first
part of the following expression, because it does not contibute to the grading.
$$\int d\sigma d\theta D\phi^\mu A_\mu(\phi)=-\int d\sigma d\theta\sum_{k\geq 0}\frac{1}{(k+1)!}\frac{k+1}{k+2}D\tilde{\phi}^\nu \tilde{\phi}^{\mu}\tilde{\phi}^{\mu_1}\dots\tilde{\phi}^{\mu_{k}}\d_{\mu_1}\dots\d_{\mu_{k}}F_{\mu\nu}(x)$$
$$-\int
d\sigma(\tilde{\psi}^\mu\psi_0^\nu+{\psi}_0^\mu\psi_0^\nu)\sum_{k\geq
  0}\frac{1}{k!}\tilde{X}^{\mu_1}\dots\tilde{X}^{\mu_k}\d_{\mu_1}\dots\d_{\mu_k}F_{\mu\nu}(x).$$
\noindent{ The}
computation is along the lines of the work by Wyllard~\cite{W} and amounts to contracting pairs of scalars using the
open-string propagator
$$D^{a,b}(\sigma)=~\alpha '\(\frac{\theta
    ^{a_ib_i}}{2\pi\alpha'}\log\(\frac{1-e ^{-\epsilon+i\sigma}}{1-e
      ^{-\epsilon-i\sigma}}\)+G^{a_ib_i}\log|1-e ^{-\epsilon+i\sigma}|^2\).$$ 
\noindent{Each} of the propagators contributing
to the regular part of the coupling comes with
two derivatives acting on two different field strengths:
$D^{a,b}\d_a\d'_b|_{x'=x}$. At order $2n$ in derivatives, the regular
part of the coupling
to $C^{(6)}$ reads   
 $$
\frac{\alpha'^n}{n!}\int_0^{2\pi}\frac{d\sigma}{2\pi}\prod_{i=1}^n\left\{\frac{\theta
    ^{a_ib_i}\d_{a_i}\d'_{b_i}}{2\pi\alpha'}\log\(\frac{1-e ^{-\epsilon+i\sigma}}{1-e
      ^{-\epsilon-i\sigma}}\)+(G^{a_ib_i}\d_{a_i}\d'_{b_i})\log|1-e ^{-\epsilon+i\sigma}|^2
 \right\}F(x)\wedge F(x').$$
\noindent{Let} us expand the product in the above expression in terms of the symmetric and antisymmetric
parts of the propagator, and remove the regulator:
$$\sum_{p=0}^nC_n^p\frac{1}{(2\pi\alpha')^p}\theta
^{a_1b_1}\dots\theta ^{a_pb_p}i ^p(\sigma-\pi)^p\times
G ^{a_{p+1}b_{p+1}}\dots G ^{a_nb_n}(\log|1- e ^{i\sigma}|^2)^{n-p}.$$
\noindent{As} we expand the gauge coupling to all orders in derivatives, we have to sum these
contributions over $n$:
$$\sum_{n\geq 0}\frac
{\alpha'^n}{n~!}\int_0^{2\pi}\(\frac{d\sigma}{2\pi}
\(\frac{\d\theta\d'}{2\pi\alpha'}i(\sigma-\pi)+ \d G\d'\log|1-e^{-i\sigma}|^2\)^n\)$$
$$= \int_0^{2\pi}\frac{d\sigma}{2\pi}\exp\(\alpha'\(\frac{\d\theta\d'}{2\pi\alpha'}i(\sigma-\pi)+
\d G\d'\log|1-e ^{-i\sigma}|^2\)\).$$
\noindent{The} last integral expression therefore equals
$$\int_0^1d\tau
|2\sin(\pi\tau)|^{2t}\exp((ia\pi)(2\tau-1))=\frac{\Gamma(1+2t)}{\Gamma(1-a+t)\Gamma(1+a+t)}.$$
\noindent{This} is a recipe for going beyond
the Seiberg--Witten limit at quadratic order in the field strength. It
confirms the prescription of~\cite{MSbeyond}, where the first order
in $G$ thereof was computed, and where larger orders were included by
requiring consistency with~\cite{star-trekIII}.

\subsection{Higher orders in the field strength}

The derivation~\cite{mezigue} of the regular part of the couplings of
larger orders in the field strength relies on symmetry factors and not on
the precise form of the propagator. It may therefore be applied here
using the full open-string propagator with derivatives included, together
with 
the following notations:
$$Q_{ij}~:=ia_{ij}(\sigma_{ij}-\epsilon(\sigma_{ij}))+t_{ij}\log\left|2\sin\(\frac{\sigma_{ij}}{2}\)\right|^2,$$
$$\sigma_{ij}~:=\sigma_i-\sigma_j,\,\,\,\,\,\,\,a_{ij}~:=\frac{\d_{i,\mu}\theta
  ^{\mu\nu}\d_{j,\nu}}{2\pi},\,\,\,\,\,\,\,t_{ij}~:=\alpha'\d_{i,\mu}G^{\mu\nu}\d_{j,\nu}.$$
\noindent{The} coupling to a mode of $C^{(10-2p)}$ is going to be expressed as the image
of $F^{p}$ by some differential operator $\tilde{\ast}_p$, so that $\tilde{\ast}_2$
is the $\ast_2{(t)}$ of ~\cite{MSbeyond}. Furthermore, if all the metric-dependent coefficients $t_{ij}$ are set to 0, the
kernel $\tilde{\ast}_p$ will reduce to the modified star-product $\ast_p$. The only change
with respect to the derivation in the Seiberg--Witten limit comes from
 the symmetric part of the propagator, which is going to insert a
 factor of $\left|2\sin\(\frac{\sigma_{ij}}{2}\)\right|^{2t_{ij}}$ in the
 integral for each pair of labels $\{i,j\}$ (with $i\neq j$, since
 extracting the regular part of the coupling prohibits self-contractions).\\

At cubic order in the field strength one can write explicitly:
 \begin{eqnarray*}  
\tilde{\ast}_3 &=& \sum_{A,B,C\geq 0}\frac{1}{A~!}\frac{1}{B~!}\frac{1}{C~!}
\int_0^{2\pi}\frac{d\sigma_1}{2\pi}\int_0^{2\pi}\frac{d\sigma_2}{2\pi}\int_0^{2\pi}\frac{d\sigma_3}{2\pi}\;
Q_{12} ^A Q_{23}^B Q_{31}^C\\
&=&\int_0^{1}d\tau_1\int_0^{1}d\tau_2\int_0^{1}d\tau_3\exp\{ia_{12}\pi(2\tau_{12}-\epsilon(\tau_{12}))
+2t_{12}\log|2\sin(\pi\tau_{12})|\\
& &
 \;\;\;\;\;\;\;\;\;\;\;\;\;\;\;\;\;\;\;\;\; \;\;\;\;\;\;\;\;\;\;\;\;\;\;\;\;\; +ia_{23}\pi(2\tau_{23}-\epsilon(\tau_{23}))+2t_{23}\log|2\sin(\pi\tau_{23})|\\
   & &\;\;\;\;\;\;\;\;\;\;\;\;\;\;\;\;\;\;\;\;\; \;\;\;\;\;\;\;\;\;\;\;\;\;\;\;\;\; +
    ia_{31}\pi(2\tau_{31}-\epsilon(\tau_{31}))+2t_{31}\log|2\sin(\pi\tau_{31})\}.
\end{eqnarray*}
{\noindent As} the symmetry factors have been shown to keep the same structure for an arbitrary
number of operators, the desired operator is seen to be the following
for any integer $p$:
$$\tilde{\ast}_p=\int_0^{1}d\tau_1\dots\int_0^{1}d\tau_p\exp\left\{\sum_{i<j}\(i\pi
  a_{ij}\(2\tau_{ij}-\epsilon(\tau_{ij})\)+2t_{ij}\log|2\sin(\pi\tau_{ij})|\) 
\right\}.$$
 Since we have disregarded from the very beginning the contact terms that
can arise from insertion of operators at the same point, we missed the
explicit counterpart of commutators that show up in the
corresponding computations in the non-commutative
description~\cite{star-trekIII} (what we derived is just the
deformation of the differential operator $\ast_p$). 
These contact terms are naturally
related to point-splitting regularization and therefore to
non-commutative gauge theory. However, the existence of
well-established Ramond--Ramond couplings in the Seiberg--Witten limit
will allow us to complete the field strength into the one of
non-commutative Yang--Mills by an educated guess, and to investigate compatibility with the 
 kernels computed above. On the other hand, the lack of explicit commutative
 treatment of these terms will restrict the range of our discussion
 of scalar couplings to deformations of the Seiberg--Witten map for
 transverse scalars.

\section{Effect on the non-commutative action}
\subsection{How to modify the smearing prescription}

It is possible to adapt the above derivation to the non-commutative
set-up, by inserting a factor of $2\sin(\pi\tau_{ab})^{2t_{ab}}$ at each of
the points at which the operators are inserted along the Wilson
line, since the full operator entering the 
coupling of degree $2p$ reads
$$\tilde{\ast}_p=\int_0^1 d\tau_1\dots\int_0^1 d\tau_p \prod_{1\leq i<j\leq p}\exp\left\{i\pi a_{ij}(2\tau_{ij}-\epsilon(\tau_{ij}))\right\}(2\sin(\pi\tau_{ij}))^{2t_{ij}}.$$
{\noindent {In}} order to see wether the couplings can be rewritten in terms of a
 smearing prescription, ordered with respect to a deformed
 star-product along a Wilson line, we are urged to find a recursive
 definition of the deformed star-products. It should be compatible the one derived
 by Liu~\cite{L} 
 between 
 the modified star-products in the Seiberg--Witten limit:
$$i\theta^{ij}\d_i\langle f_1,\dots,f_{p},\d_j
f_{p+1}\rangle_{\ast_{p+1}}
=\sum_{i=1}^{p}\langle f_1,\dots,[f_i,f_{p+1}],\dots,f_{p}\rangle_{\ast_{p}}.$$
\noindent{where} the commutator is understood with respect to the
star-product. As noted in~\cite{MSbeyond}, the commutator can still
be expressed in terms of $\ast_2$ after deformation:
$$i\theta ^{ij}\langle\d_i f,\d_j
g\rangle_{\ast_2(t)}=[f,g]_{\ast(t)},$$
\noindent{where $\ast(t)$ is the deformed version of the star-product
  defined by the prescription of integration along an open Wilson line
 for two observables.
 Therefore we are inclined to look for a deformed version of the recursive formula
using derivatives for some of the arguments. 
Let us onsider the following expression
$$i\theta^{ij}\d_i\langle f_1,\dots,f_{p},\d_j
f_{p+1}\rangle_{\tilde{\ast}_{p+1}},$$
and show how the recursion is organized for one of the
terms in the above derivative. It is of the general form of multiple
convolutions (denoted by $\circ$) between operators $O_i$ smeared along a line, where the
kernel $K$
is translation invariant.
\begin{eqnarray*}
&&i\theta ^{ij}\d_iO_1\circ \d_jO_2\circ\dots\circ
O_{p+1}\\
&=&i\theta ^{ij}\int_0^1 d\tau_1\dots \int_0^1 d\tau_{p}\, \d_i O_1(0)
K(\tau_1)\d_j O_2(\tau_1)K(\tau_2)O_3(\tau_1+\tau_2)K(\tau_3)O_4(\tau_1+\tau_2+\tau_3)\dots\\
&&\dots  K(\tau_{p-1})O_{p}(\tau_1+\dots+\tau_{p-1})K(\tau_p)O_{p+1}(\tau_1+\dots+\tau_p)
\delta(\tau_1+\dots+\tau_p-1) \\  
 &=&\(i\theta ^{ij}\int d\tau_1 \d_i O_1(0)
 K(\tau_1)\d_j O_2(\tau_1)\)\\
&&\times \int_0^1 d\tau_2\dots \int_0^1 d\tau_{p}\,
K(\tau_2) O_3(1-\tau_3-\dots-\tau_p) K(\tau_3)\dots O_{p}(1-\tau_p)K(\tau_p)O_{p+1}(1)\\
&=&\int_0^1 d\tau_2\dots \int_0^1 d\tau_{p}[O_1,O_2]_{\ast(t_{12})}(0)
K(\tau_2)O_3(\tau_2)K(\tau_3) O_4(\tau_2+\tau_3)\dots\\
&&\dots
O_{p}(\tau_2+\dots+\tau_{p-1})K(\tau_p)O_{p+1}(\tau_2+\dots+\tau_p)\delta(\tau_2+\dots+\tau_p-1)\\
&=&[O_1,O_2]_{\ast(t_{12})}\circ O_3\circ\dots\circ O_{p}\circ{O_{p+1}}.
\end{eqnarray*}
\noindent{This} allows us to open up one more integration interval over an intermediate
time and to write the recursive definition of the deformed smearing prescription 
$$i\theta^{ij}\d_i\langle f_1,\dots,f_{p},\d_j
f_{p+1}\rangle_{ \tilde{\ast}_{p+1}
}=\sum_{i=1}^{p}\langle f_1,\dots,[f_i,f_{p+1}]_{\ast(t_{i,{p+1}})},\dots,f_{p}\rangle_{\tilde{\ast}_{p}}.$$
\noindent{Let} us write the third rank differential operator explicitly: 
$$\tilde{\ast}_3=\frac{a_{32}}{a_{31}+a_{32}}\frac{\Gamma\( 1+2t_{32}\)}{\Gamma\(1-a_{32}+t_{32}
  \)\Gamma\(1+a_{32}+t_{32} \)}\frac{\Gamma\(1+2t_{12}+2t_{13}
  \)}{\Gamma\(1-a_{12}-a_{13}+t_{12}+t_{13}
  \)\Gamma\(1+a_{12}+a_{13}+t_{12}+t_{13}  \)}$$
$$+\frac{a_{31}}{a_{31}+a_{32}}\frac{\Gamma\( 1+2t_{31}\)}{\Gamma\(1-a_{31}+t_{31}
  \)\Gamma\(1+a_{31}+t_{31} \)}\frac{\Gamma\(1+2t_{32}+2t_{12}
  \)}{\Gamma\(1-a_{32}-a_{12}+t_{32}+t_{12}
  \)\Gamma\(1+a_{32}+a_{12}+t_{32}+t_{12}  \)},$$
\noindent{whose} Seiberg--Witten limit is recognized as $\ast_3$:
$${\ast}_3=\frac{\sin\(\pi
  a_{32}\)\sin\(\pi(a_{12}+a_{13})\)}{\pi(a_{31}+a_{32})\pi(a_{12}+a_{13})}+
\frac{\sin\(\pi
  a_{31}\)\sin\(\pi(a_{32}+a_{12})\)}{\pi(a_{31}+a_{32})\pi(a_{32}+a_{12})}
=\lim_{\alpha'\rightarrow 0} \tilde{\ast}_3.$$

\subsection{Deformed non-commutative gauge transformations}
The previous investigation of deformed star-products at larger degree
allows to derive the deformation of non-commutative field strength and
gauge transformation required to ensure gauge invariance of the
deformed smeared expression. These are as announced in~\cite{MSbeyond}:  
$$\hat{F}_{ij}=\d_i\hat{A}_j-\d_j\hat{A}_i-i[\hat{A}_i,\hat{A}_j]_{\ast(t)},$$
$$\delta\hat{A}_i=\d_i\hat{\lambda}-i[\hat{A}_i,\hat{\lambda}]_{\ast(t)}.$$
\noindent{The} recursive formula was the custodian of gauge invariance in the
Seiberg--Witten limit. Extending our prescriptions to larger orders in
the gauge potentials by expanding the deformed Wilson line, we see
that the deformed smeared prescription plays exactly the same role.
 The Ramond--Ramond couplings $Q(k)$ are given, for some mode $k$, by a
 smeared integral along a straight open Wilson line $W_k$ (of
 extension $\theta ^{\mu\nu} k_\nu$), of operators $O_i$
 of the form $(\theta -\theta\hat{F}\theta)^{\mu\nu}$, transforming as 
$$O_i\mapsto -i[O_i,\hat{\lambda}]_{\tilde{\ast}},$$
\noindent{so} that the gauge invariance of the couplings can be checked on an
 expansion in terms of the gauge potential, using the
recursive definition of the deformed star-products:  
  $$Q(k)=\sum_{m\geq 0} Q_m(k),$$
$$Q_m=\frac{1}{m~!}(\theta\d)^{\mu_1}\dots(\theta\d)^{\mu_m}\langle
O_1,\dots,O_p,\hat{A}_{\mu_1},\dots,\hat{A}_{\mu_m} \rangle_{\tilde{\ast}_{p+m}}.$$
\noindent{The gauge} variation of one of the terms in the above
  expansion reads 
\begin{eqnarray*}
\delta
Q_m&=&-\frac{i}{m~!}(\theta\d)^{\mu_1}\dots(\theta\d)^{\mu_m}\sum_{i=1}^p\langle
O_1,\dots,[\hat{\lambda},O_i]_{\tilde{\ast}},\dots,
  O_p,\hat{A}_{\mu_1},\dots,\hat{A}_{\mu_m}
  \rangle_{\tilde{\ast}_{p+m} }\\
&&-\frac{i}{m~!}(\theta\d)^{\mu_1}\dots(\theta\d)^{\mu_m}\sum_{i=1}^m\langle
O_1,\dots,
  O_p,\hat{A}_{\mu_1},\dots,[\hat{\lambda},\hat{A}_{\mu_i}]_{\tilde{\ast}},\dots,\hat{A}_{\mu_m} \rangle_{\tilde{\ast}_{p+m} } \\
&&+\frac{1}{(m-1)~!}(\theta\d)^{\mu_1}\dots(\theta\d)^{\mu_{m}}\langle
O_1,\dots,O_p,\hat{A}_{\mu_1},\dots,\hat{A}_{\mu_{m-1}},\d_{\mu_{m}}\hat{\lambda}
\rangle_{\tilde{\ast}_{p+m}},
\end{eqnarray*} 
\noindent{so} that the gauge variation of the field strengths in $Q_m$ is
compensated by the gauge variation of the gauge potentials in
$Q_{m+1}$. The quantity $Q(k)$ is therefore gauge invariant, and the
deformed smearing prescription is consistent in the non-commutative
set-up, provided the commutators of non-commutative Yang--Mills theory
are also deformed. Furthermore, we may infer deformations of the Seiberg--Witten map for the
transverse scalars, by considering a lower-dimensional brane and
identifying the coefficients of the transverse momentum of the Wilson
line in both
descriptions. This results in the substitution of deformed
star-products to the usual ones in the corresponding result of Mukhi
and Suryanarayana~\cite{CSterms}.\\

The family of differential operators (and the deformed gauge theory) we have just worked out induce
deformations of the expressions of the form 
$$ \int dx \;L_{\ast}\(\sqrt{\det(1-\theta
  \hat{F})}\(\hat{F}\frac{1}{1-\theta \hat{F}}\)^p W_k(x)\)\ast e ^{ikx},$$  
 by replacing star-products of various ranks (and field
strengths) with their deformations.  A few more 
 terms in the commutative set-up can back this
 proposal. During the computation on a commutative space, we ignored terms
 that involved other tensor structures than derivatives of 
$F\wedge F$. But an important class of such terms are predicted by the
 modified smearing prescription, since the Wilson line not
 only gives rise to an ordering of the observables, but can be
 expanded, generating forms of degree four and of cubic order in the
 gauge field, even once expressed back in the commmutative
 language. The Seiberg--Witten limit of these forms has been worked
 out in~\cite{DMS}. Terms from the expansion of the open Wilson lines that are cubic in the field srength arise through the four-form
$$\frac{1}{2}\theta ^{\mu\nu}\partial_\nu\langle\hat{A}_\mu,\hat{F}_{\alpha\beta}, \hat{F}_{\gamma\delta}\rangle_{\tilde{\ast}_3}.$$
The commutative counterpart~\cite{W} (at low order in derivatives) of these terms with all the form
 indices carried by two field strengths is in the four-derivative
 four-form term 
$$\theta ^{\mu\nu}\theta ^{\rho\kappa} \theta ^{\sigma\tau} F_{\rho\mu}\d_\sigma\d_\nu F_{\alpha\beta} \d_\kappa\d_\tau
F_{\gamma\delta},$$
and modifications thereof beyond the Seiberg--Witten limit. The relevant modifications are quadratic in the
metric, since the two-forms commute with each other, forcing the two differential operators
acting on them to have the same symmetry. The relevant tensor structure is therefore
as follows:
 $$\theta ^{\mu\nu} F_{\mu\rho}  G^{\rho\kappa} G ^{\sigma\tau}\d_\nu\d_\sigma F_{\alpha\beta}\d_\kappa\d_\tau
 F_{\gamma\delta}.$$
Now, to be consistent on the non-commutative side, we must take into account
  the contribution from the factor  $\sqrt{\det(1-\theta
  \hat{F})}$ to the four-form coupling with the relevant index structure
$$-\frac{1}{4} \theta ^{\mu\nu}\langle \hat{F}_{\nu\mu}, \hat{F}_{\alpha\beta},\hat{F}_{\gamma\delta}\rangle_{\tilde{\ast}_3},$$
and the cubic part of the Seiberg--Witten image of
$\langle\hat{F}_{\alpha\beta},\hat{F}_{\gamma\delta}\rangle_{\tilde{\ast}_2}$,
which read
 $$-\theta ^{\mu\nu} \langle\(\langle \hat{A}_\mu ,\partial_\nu
\hat{F}_{\alpha\beta}\rangle_{\tilde{\ast}_2}-
\langle \hat{F}_{\alpha\mu},\hat{F}_{\beta\nu}\rangle_{\tilde{\ast}_2}\), \hat{F}_{\gamma\delta}\rangle_{\tilde{\ast}_2}.$$
We may note that this expression includes deformations due the ones of
{\it gauge transformations}, while the new terms from the open Wilson line are
direct consequences of the deformation of the {\it star-products}. The
two levels of our previous discussion are therefore tied together.  
Let us trace the modifications of the commutative terms to the
deformations of the open Wilson line. 
They can only come from the terms in the open
Wilson line where the overall derivative acts on one of the field
strengths. The first contribution of the differential operator $t$ is 
 quadratic, as could be awaited: 
$$\tilde{\ast}_2-\ast_2=\frac{\pi ^2}{3} t ^2 +o(t ^2).$$
$$\tilde{\ast}_3-\ast_3=\frac{\pi ^2}{6} (t_{13}^2+(t_{12}+ t_{23})^2) +o(t ^2).$$
In order to make a field strength out of the gauge field $\hat{A}_i$,
one has to use derivatives under the disguise of $t^2$. This produces
terms where each of the two-forms bears a pair of derivatives, and where one
of these pairs contains two derivatives contracted with inverse
open-string metrics:
$$\theta ^{\mu\nu}\partial_\kappa \hat{A}_\mu G^{\tau\sigma}\partial_\tau\partial_\nu
\hat{F}_{\alpha\beta} G^{\kappa\rho} \partial_\rho\partial_\sigma\hat{F}_{\gamma\delta}.$$
The gauge-invariant completion comes from the expansion of the
external $\tilde{\ast}_2$ in the {\hbox{Seiberg--Witten}} map. The index
structure of the commutative candidate is recognized after rearranging, and the removal
of hats is consistent with our cubic prescription.

\section{Conclusion}

In order to obtain results beyond the Seiberg--Witten limit, the full open-string propagator has been taken into account in the
computation of the Ramond--Ramond couplings for small $U(1)$ 
field strength
 at all orders in derivatives. The resulting differential operators
 acting on powers of the field strength are deformations of the
 modified star-products previously derived in the Seiberg--Witten
 limit. The expression of $\tilde{\ast}_2$ is consistent with known string
 amplitudes. Moreover, the recursive definition
 of the modified star-products enables us to express any of them as rational functions of differential operators containing open-string parameters, and Gamma functions thereof.  
The results can be reformulated in the
 non-commutative set-up in terms of a deformed smearing prescription
  along an open Wilson line. This confirms earlier proposals at all
 orders in the open-string metric at disk level, and extends them to
 couplings of higher degrees. Furthermore, the deformation of the commutators
   induced in non-commutative {\hbox {Yang--Mills}} theory by
 $\tilde{\ast}_2$ has been shown to lead to gauge-invariant
 couplings.\\

\noindent{{{ \Large \bf Acknowledgements}}}\\ 

I have benefited from conversations with R. Minasian, D. Orlando and P. Vanhove,
and from correspondence with S. Mukhi. It
is a pleasure to thank the organizers of the Les Houches Winter
School {\it Frontiers in Number Theory, Physics, and Geometry} (HPCF 2002-00334), during which
part of this work was done.


\begin{thebibliography}{beyond}


 \bibitem {CH} C.-S. Chu and P.-M. Ho, \emph{Non-Commutative Open String and D-brane}, Nucl. Phys. {\bf B550} (1999) 151-168,  {\ttfamily hep-th/9812219}. 

\bibitem {SW} N. Seiberg and E. Witten, \emph{String Theory and Non-Commutative Geometry}, JHEP {\bf 9909} (1999) 032, {\ttfamily hep-th/9908142}.


\bibitem {Cornalba} L. Cornalba, \emph{D-branes Physics and Non-Commutative
    Yang--Mills Theory}, Adv. Theor. Math. Phys. {\bf 4} (2000)
  271-281, {\ttfamily hep-th/9909081}.

\bibitem  {L} H. Liu, \emph{$\ast$-Trek II: $\ast_n$ Operations, Open Wilson Lines and the Seiberg--Witten Map}, Nucl. Phys. {\bf B614} (2001) 305-329, {\ttfamily hep-th/0011125}.

  

\bibitem {OO} Y. Okawa and H. Ooguri, \emph{An Exact Solution to Seiberg--Witten
  Equation of {\mbox{Non-Commutative}} Gauge Theory}, Phys. Rev. {\bf D64} (2001) 046009, {\ttfamily hep-th/0104036}.




\bibitem {CSterms} S. Mukhi and N.V. Suryanarayana, \emph{Gauge-Invariant Couplings of Non-Commutative Branes to Ramond--Ramond Backgrounds}, JHEP {\bf{0105}} (2001) 023, {\ttfamily hep-th/0104045}.

   
\bibitem  {LM} H. Liu and J. Michelson, \emph{Ramond--Ramond Couplings of Non-Commutative {\mbox{D-Branes,}}} Phys. Lett. {\bf B518} (2001) 143-152, {\ttfamily hep-th/0104139}.





\bibitem {DMS} S.R. Das, S. Mukhi and N.V. Suryanarayana, \emph{ Derivative Corrections from {\mbox{Non-Commutativity}}}, JHEP {\bf 0108} (2001) 039, {\ttfamily hep-th/0106024}.






\bibitem{Rey1} S.-J. Rey and R. von Unge, \emph{S-Duality, Non-Critical
    Open String and Non-Commutative Gauge Theory}, Phys. Lett. {\bf B499}
    (2001) 215-222, {\ttfamily hep-th/0007089}.



\bibitem{Rey2} S. Das and S.-J. Rey, \emph{Open Wilson Lines in Non-Commutative Gauge Theory and Tomography of Holographic
      Dual Supergravity}, Nucl. Phys. {\bf B590} (2000) 453-470, {\ttfamily hep-th/0008042}.




\bibitem {Gross}  D.J. Gross, A. Hashimoto and N. Itzhaki, \emph{ Observables of Non-Commutative Gauge {\mbox{Theories}}}, Adv. Theor. Math. Phys. {\bf 4} (2000) 893-928, {\ttfamily hep-th/0008075}.



\bibitem {star-trekI} H. Liu and J. Michelson, \emph{$\ast$-TREK: The
    One-Loop N=4 Non-Commutative SYM Action}, Nucl. Phys. {\bf B614}
  (2001) 279-304, {\ttfamily hep-th/0008205}.
 



\bibitem{MW} T. Mehen and M. Wise, \emph{Generalized $\ast$-Products,
    Wilson Lines and the Solution of the Seiberg-Witten Equations},
  JHEP {\bf 0012} (2000) 008, {\ttfamily hep-th/0010204}.



\bibitem{DT} S.R. Das and S.P. Trivedi, \emph{Supergravity Couplings to Non-Commutative Branes, Open Wilson Lines
and Generalised Star-Products},
  JHEP {\bf 0102} (2001) 046, {\ttfamily hep-th/0011131}.



\bibitem{ReyHouches} S.-J. Rey, \emph{Exact Answers to Approximate Questions: Non-Commutative Dipoles, Open
Wilson Lines, and UV-IR Duality}, Lecture given at Les Houches
Summer School, Session 76, {\it Unity of
Fundamental Physics: Gravity, Gauge Theory and Strings}, {\ttfamily hep-th/0207108}.

\bibitem {M} S. Mukhi, \emph{ Star-Products from Commutative String Theory}, Pramana {\bf 58} (2002) 21-26, {\ttfamily hep-th/0108072}.



\bibitem {mezigue} P. Grange, \emph{Derivative Corrections from Boundary State Computations}, Nucl. Phys. {\bf
  B649} (2003) 297-311, {\ttfamily hep-th/0207211}.


  

\bibitem {MSbeyond} S. Mukhi and N.V. Suryanarayana, \emph{Open-String
    Actions and Noncommutativity Beyond the Large-B Limit},  JHEP {\bf
    0211} (2002) 002, {\ttfamily hep-th/0208203}.


  

\bibitem {W} N. Wyllard, \emph{ Derivative Corrections to D-brane Actions with Constant Background Fields}, Nucl. Phys. {\bf B598} (2001) 247-275, {\ttfamily hep-th/0008125}.



\bibitem{star-trekIII} H. Liu and J. Michelson, 
 \emph{$\ast$-Trek III: The Search for Ramond-Ramond Couplings},
  Nucl. Phys. {\bf B614} (2001) 330-366, {\ttfamily hep-th/0107172}.


\end{thebibliography}
\end{document}